\documentclass[12pt]{article}

\usepackage{amssymb}

\newcommand{\dd}{\partial}
\newcommand{\de}{\delta}
\newcommand{\m}{\mu}
\newcommand{\n}{\nu}
\newcommand{\ls}{\left(}
\newcommand{\rs}{\right)}
\newcommand{\lks}{\left[}
\newcommand{\rks}{\right]}
\newcommand{\po}{{\Pi_{\!\!\bot}}}
\newcommand{\al}{\alpha}
\newcommand{\be}{\beta}
\newcommand{\dz}{\zeta}
\newcommand{\lc}{{\cal L}}
\newcommand{\ff}{\varphi}
\newcommand{\ka}{\varkappa}

\newcommand{\disn}[2]{$$\displaylines{\refstepcounter{equation}%
            \label{#1}\hskip 1em minus 1em #2\hfilneg}$$}
\newcommand{\nom}{\hfil\hskip 1em minus 1em (\theequation)}

\newcommand{\nss}{\hfill\cr\hfill}

\makeatletter
\renewcommand{\section}{\@startsection{section}{1}{0pt}%
          {3.5ex plus 1ex minus .2ex}{2.3ex plus .2ex}{\noindent\hfil\bf}}
\makeatother

\begin{document}

\title{Gravity as a field theory in flat space-time}
\author{
S.A.~Paston\thanks{E-mail: paston@pobox.spbu.ru}\\
{\it Saint Petersburg State University, St.-Petersburg, Russia}
}
\date{\vskip 15mm}
\maketitle

\begin{abstract}
We propose a formulation of gravity theory in the form of a field theory in a flat space-time with a number
of dimensions greater than four. Configurations of the field under consideration describe the splitting of
this space-time into a system of mutually noninteracting four-dimensional surfaces. Each of these surfaces
can be considered our four-dimensional space-time. If the theory equations of motion are satisfied, then
each surface satisfies the Regge-Teitelboim equations, whose solutions, in particular, are solutions of the
Einstein equations. Matter fields then satisfy the standard equations, and their excitations propagate only
along the surfaces. The formulation of the gravity theory under consideration could be useful in attempts
to quantize it.
\end{abstract}

\newpage
\section{Introduction}
Einstein's General Relativity is the commonly accepted theory of gravity. If we stay within classical
(i.~e., nonquantum) physics, then this theory describes established observational facts well. But attempts to
construct a closed theory of quantum gravity meet very serious obstacles. It must be admitted that despite
the large number of papers devoted to various approaches to solving this problem, a commonly accepted
closed theory of quantum gravity is still lacking. The best-known reason for this is that quantizing gravity
in terms of the space-time metric $g_{\m\n}(x)$ results in a theory that is perturbatively nonrenormalizable with
respect to variations of the metric over a flat background (see \cite{carlip} and the references therein). We note that
in this approach, the procedure for quantizing gravity is constructed by analogy with quantizing a field in
a flat space-time.

But nonrenormalizability is not the only problem that appears when quantizing gravity. We encounter
other problems, which are possibly even deeper, related to the fact that gravity is determined by space-time
properties, and we intend to quantize just this space-time. The positive experience of quantizing theories,
such as quantum electrodynamics, quantum chromodynamics, etc., in a flat space-time is then of little help.
We mention one such problem, the problem of formulating the causality principle.

In flat space-time, we postulate that field operators located at spacelike-separated points must commute.
This postulate results in the canonical commutation relations underlying canonical quantization. In
the case of gravity, whether two points are separated by a spacelike or a timelike interval is determined by
the metric $g_{\m\n}(x)$, which becomes an operator after quantization, and we are hence unable to provide a
definite answer. As the result, we cannot consistently define how values of the field $g_{\m\n}(x)$ taken at different
points commute, and applying the customary quantization scheme seems less justified.

A discussion of other problems that appear when quantizing gravity, in particular, the very important
problem of choosing the time can be found in \cite{carlip}, where references to the original papers on the subject can
also be found. In fact, all these problems arise because we try to implement a quantization procedure that
worked well when applied to field theories in the flat space-time to the case where the dynamical variables
are the geometric properties of the space-time, i.~e., where we must quantize the space-time structure itself.

A possible way to resolve the abovementioned problems could amount to passing from a direct attempt
to quantize gravity (e.g., acting by analogy with the quantization of electrodynamics) in the same terms
with which we describe it on the classical level (in terms of the metric tensor $g_{\m\n}(x)$) to constructing a
quantum theory from which the gravity theory would follow in some sense and, perhaps, in some limit. The
superstring theory and the loop theory of gravity are premier examples of such an approach to constructing
a quantum theory of gravity. A brief discussion of these theories from the standpoint of constructing a
quantum theory of gravity and also a description of the problems that so far prevent these theories from
being considered a comprehensive solution of the problem of such a construction can be found in the review
cited above \cite{carlip}. We note that both these examples are not a quantized theory of a field in a flat space-time.

Because the quantum theories of all interactions except the gravitational one were successfully constructed
in the framework of such theories, it seems plausible to obtain gravity as just a quantum field
theory in a flat space-time. It then seems reasonable to first formulate a classical theory of gravity as
a classical field theory in a flat space-time and subsequently use the standard quantization procedure to
obtain a quantum theory of gravity free of the above problem.

Here, we propose a formulation of the theory of gravity as a field theory in the $N$-dimensional ($N>4$)
flat space-time $R^{1,N-1}$ with one timelike and $N-1$ spacelike dimensions. This formulation develops the
theory of embedding proposed by T.~Regge and C.~Teitelboim in \cite{regge}, where gravity is described by analogy with
string theory. The theory of embedding assumes that our space-time is a four-dimensional surface embedded
in a flat space-time $R^{1,N-1}$. The independent variable describing gravity is then the function $y^a(x^\m)$ of
embedding of this surface in the ambient space (here and hereafter, $a,b,\ldots=0,1,2,\dots,N-1$ and $y^a$ are
the Lorentz coordinates in the space $R^{1,N-1}$). The metric is expressed in terms of this coordinate as
 \disn{v1}{
g_{\m\n}=\eta_{ab}\,\dd_\m y^a\, \dd_\n y^b,
\nom}
where $\eta_{ab}$ is the pseudo-Euclidean metric of the space $R^{1,N-1}$.

Choosing the standard Einstein-Hilbert expression as the embedding theory action, we find that the
equations of motion are the Regge-Teitelboim equations, which are more general than the Einstein equations.
That is, in addition to all the solutions of the Einstein equation, these equations also contain other
(so-called extra) solutions. These solutions can be eliminated by introducing additional constraints into
the theory as previously proposed in \cite{regge} and investigated in detail in \cite{sbshk05}, \cite{tmf07} for $N=10$. We note that this
value of $N$ is distinguished because in accordance with the Janet and Cartan theorem \cite{gane}, \cite{kart} (see, e.g.,
Remark 18 in \cite{kobno2}), an arbitrary four-dimensional Riemannian space can be locally isometrically embedded
in just the ten-dimensional space. The correct form of the algebra of first-class constraints, which appears in
the canonical formalism for the embedding theory when imposing additional constraints eliminating extra
solutions, was found and investigated in \cite{tmf07}, \cite{ijtp10}. In what follows, we can use an analogous approach for
eliminating extra solutions in the theory proposed here because this theory also admits extra solutions. We
note that instead of eliminating extra solutions, another way is possible: we can investigate these solutions
trying to determine their physical content possibly under imposing proper boundary conditions at infinity.
This problem and also the problem of choosing the most appropriate value of the dimension $N$ have not
yet been investigated.

But the theory of embedding, being a theory in the flat space $R^{1,N-1}$, is not a field theory, and the
above quantization problems that appear in the standard description of gravity are to a large extent also
intrinsic to this theory. For example, the causality problem remains: the function $y^a(x)$ when quantized
becomes an operator whose commutation relations are again difficult to determine because this operator
depends on the coordinates $x^\m$ of points on a surface whose interval is determined by the metric related to
the very same operator by formula (\ref{v1}).

We can consider the embedding theory as the theory of one three-dimen\-sio\-nal brane (in spatial directions),
which as time passes describes a four-dimensional surface embedded in a flat space-time. We can
then indicate an analogy between this brane and the mechanics of a point mass, which is the one-particle
theory describing a one-dimensional worldline in the Minkowski space as time passes. The difference is
only due to the dimension and due to the fact that the action is not just a volume but the integral of the
scalar curvature. We can pass to the field theory if we pass from a single three-dimensional brane to the
medium composed of branes that fill the whole ambient flat space analogously to how we can consider a
medium composed of particles rather than a single particle. In other words, we consider the field theory
describing many Regge-Teitelboim surfaces filling the whole space $R^{1,N-1}$, i.~e., passing through each point
of this space.

We assume that the Regge-Teitelboim surfaces (denoted by $W$) do not intersect, i.~e., that we have
exactly one surface passing through a point of the space, and that different surfaces do not interact (or
almost do not interact) with each other. We can say that we make "splitting" the flat space $R^{1,N-1}$ into a
system of surfaces $W$. For brevity, we call the field theory describing such a system of surfaces the
splitting theory. We also assume that in addition to the field describing the system of surfaces, other fields can be
present in the space $R^{1,N-1}$, namely, matter fields. Although these fields are understood as being defined
in the ambient space, we try to organize their interaction to produce a theory such that all interactions
would propagate only along the four-dimensional curved surfaces $W$ whose geometry would correspond to
solutions of the Einstein equations. Then any of the surfaces $W$ can be considered our four-dimensional
space-time. This paper is devoted to constructing such a splitting theory.

In Sec. 2, we propose an explicit description of the system of surfaces $W$ using one field and demonstrate
how we can define the scalar curvature of each surface in terms of this field. We construct the theory action
that ensures the absence of interactions between different surfaces $W$ in Sec. 3 and derive the equations
of motion corresponding to this action in Sec. 4. We obtain the standard equations of motion for the
matter fields and the Regge-Teitelboim equations for the gravitation field. In particular, this means that
all excitations propagate only along the surfaces $W$ and all solutions of the Einstein equation are solutions
of the obtained theory.

\section{The splitting theory}
In the flat $N$-dimensional Minkowski space $R^{1,N-1}$, we consider the theory of the real $N-4$-component
field $z^A(y^a)$, where $y^a$ are the Lorentz coordinates in the space $R^{1,N-1}$, $a=0,\ldots,N-1$, $A=1,\ldots,N-4$.
We do not rigidly fix the signature of the space $R^{1,N-1}$ assuming only that it has the form $\{\dz,-\dz,\ldots,-\dz\}$,
where $\dz=\pm1$, i.~e., that the timelike direction is unique.

We associate a splitting of the space $R^{1,N-1}$ into a system of surfaces $z^A(y)=const$, i.~e., surfaces
of the constant field value, with each field configuration $z^A(y^a)$. It is clear that apart from degenerate
situations, the surfaces $W$ are four-dimensional. We assume that the field theory is invariant under the
field transformation
\disn{1}{
z^A(y)\longrightarrow z'^A(y)=f^A(z(y)),
\nom}
where $f^A(z)$ is an arbitrary function. This symmetry indicates that physical degrees of freedom depend
only on how we make splitting the space $R^{1,N-1}$ into a system of surfaces, not on the actual values of the field
$z^A$ on each of the surfaces.

We assume that one of the surfaces $z^A(y)=const$ (any of them) is our space-time. Then the intrinsic
geometry of our space-time, which determines the gravitational interaction, is determined by the shape of
the surface and therefore by the configuration of the field $z^A(y)$, and transformation (\ref{1}) does not change
this geometry. To write the majority of characteristics of the intrinsic geometry, the metric for example, we
must introduce a coordinate system on $W$. But all the invariant characteristics are uniquely determined
by the shape of the surface $W$ and hence by the field configuration $z^A(y)$. Describing the system in terms
of this field, we therefore obtain an invariant description of gravity that does not rely on a particular
coordinate system.

We introduce the notation
\disn{2}{
\frac{\dd}{\dd y^a}z^A(y)\equiv \dd_a z^A\equiv v^A_a.
\nom}
We note that $v^A_a$ at fixed values of $A$ is a set of vectors normal to $W$. The space of values of the function
$z^A(y)$ is an $(N-4)$-dimensional manifold, denoted here by $Z$. Transformation (\ref{1}) is a change of coordinates
on this manifold, i.~e., it is a "general covariant" transformation in the space of field values (we customarily
call a general covariant transformation a change of coordinates in the space of field arguments). We can
consider $v^A_a$ a map (which is one to one in the general position case) between the tangent space at a given
point of the manifold $Z$ (understood as the linear space of differentials of coordinates on $Z$) and the space
orthogonal to $W$ at this point. It is important that all the points of a given surface $W$ correspond to the
same point in $Z$ and spaces orthogonal to the same surface $W$ at different points are mapped to $Z$ in the
same tangent space.

Because $R^{1,N-1}$ is flat, we have the metric $\eta_{ab}={\rm diag}(\dz,-\dz,\ldots,-\dz)$
in this space. Using this metric,
we can construct the quantity
\disn{3}{
w^{AB}(y)=v^A_a v^B_b\eta^{ab},
\nom}
which can be considered a metric in $Z$. But it must be remembered that this metric depends on $y^a$ and
hence not only on the point in $Z$ but also on the position of the surface $W$ corresponding to this point.

We assume that the surfaces $W$ always contain a timelike direction. We formulate this condition as
the condition of the sign-definiteness of the matrix $w^{AB}$ (the sign depends on the choice of the signature):
\disn{9}{
\dz w^{AB} s_As_B<0\qquad \forall\,\,\, s_A.
\nom}

We define the quantity $w_{AB}(y)$ as the matrix inverse to (\ref{3})
and introduce the notation for the determinant $w\equiv\det(w^{AB})$.
We raise and lower indices of the type $A,B,\ldots$ using wab and $w_{AB},w^{AB}$. For example,
it is convenient to introduce the quantity inverse to $v^A$ in a certain sense,
\disn{6}{
v_A^a=w_{AB}v^A_b\eta^{ba},
\nom}
for which the relations
\disn{7}{
v^A_a v^a_B=\de^A_B,\qquad v^A_a v^b_A=\po^b_a,
\nom}
hold, where $\po^b_a$ is the projection operator on the space orthogonal to $W$ at the given point. We can easily
write the corresponding operator of projection on the tangent space:
\disn{8}{
\Pi^b_a=\de^b_a-\po^b_a.
\nom}
We note that the projection operators $\Pi^b_a$ and $\po^b_a$,
as is easily seen, are invariant under transformation (\ref{1}).

After expressing the tangent and orthogonal projection operators for $W$ in terms of the independent
variable $z^A(y)$, we can express the scalar curvature $R$ of the surface in terms of this variable (see the
explicit formula below). To operate with nonscalar characteristics of the intrinsic geometry of the surface,
we introduce a temporary coordinate system $x^\m$ on each surface $W$ (here $\m=0,1,2,3$). Because we
define $x^\m$ on every surface, we obtain a function $x^\m(y)$. We can consider the set $\{z^A(y),x^\m(y)\}$ as curvilinear
coordinates in the flat space $R^{1,N-1}$.

We use the formula for the second fundamental form on a surface defined by the function $y^a(x)$ of
embedding into a flat ambient space (an exposition of the embedding theory formalism can be found in \cite{tmf07}
and in more detail in \cite{piyf06}):
\disn{8.1}{
b^a_{\m\n}=e^b_\m\dd_\n\Pi^a_b,
\nom}
where $\dd_\n\equiv\dd/\dd x^\n$ and $e^b_\m=\dd_\m y^b$.
In the case considered here, the projection operator $\Pi^a_b$ can be considered
not only a function of the coordinates $x^\m$ but also a function of the point $y^a$ of the space $R^{1,N-1}$. We can
therefore write
\disn{11}{
\dd_\n\Pi^a_b=\frac{\dd}{\dd x^\n}\Pi^a_b=\frac{\dd y^e}{\dd x^\n}\frac{\dd}{\dd y^e}\Pi^a_b=e^e_\n\dd_e\Pi^a_b.
\nom}
Using this relation, we "transfer" the second fundamental form of a surface into the ambient space with
respect to all its indices:
\disn{8.2}{
b^a_{cd}\equiv e^\m_c e^\n_d\, b^a_{\m\n}=\Pi^b_c\Pi^e_d\,\dd_e\Pi^a_b=\Pi^b_c\,\bar\dd_d\Pi^a_b,
\nom}
where $e^\m_c=\eta_{ce}g^{\m\n}e_\n^e$ and we introduce the notation for the tangent derivative
\disn{13}{
\bar\dd_d\equiv\Pi^e_d\dd_e=e^\m_d\dd_\m.
\nom}
Using the projection operator properties (see \cite{tmf07}, \cite{piyf06}) and relation (\ref{7}), we can continue formula (\ref{8.2}):
\disn{13.1}{
b^a_{cd}=-\Pi^b_c\,\bar\dd_d\po^a_b=-\Pi^b_c\ls\bar\dd_d v_b^A\rs v_A^a=
-\Pi^b_c\Pi^e_d\ls\dd_e\dd_b z^A\rs v_A^a.
\nom}

We express the Riemann tensor in terms of the second fundamental form as
\disn{8.3}{
R_{\al\be\m\n}=\lks \eta_{ef}\, b^e_{\al\m}\,b^f_{\be\n}\rks_{\m\n},
\nom}
where we imply the anti-symmetrization with respect to the indices $\m$ and $\n$: $[O_{\m\n}]_{\m\n}=O_{\m\n}-O_{\n\m}$. We
can again transfer it to the ambient space with respect to all its indices:
\disn{11.1}{
R_{abcd}\equiv e^\al_a e^\be_b e^\m_c e^\n_d\, R_{\al\be\m\n}=
\Pi_a^e\Pi_b^f\Pi_c^g\Pi_d^h\lks\ls\dd_e\dd_g z^A\rs w_{AB}\ls\dd_f\dd_h z^B\rs\rks_{gh}.
\nom}
The coordinates $x^\m$ already do not enter expressions (\ref{13.1}) and (\ref{11.1}),
and we can therefore use these expressions
without referring to systems of coordinates on the surfaces $W$. We note that the quantity $R_{abcd}$ is "tangent"
with respect to all its indices, i.~e., it satisfies identities of the type $\po^a_e R_{abcd}=0$.

We can now write the expressions for the Ricci tensor, for the scalar curvature, and for the Einstein
tensor without referring to coordinates on $W$:
\disn{12}{
R_{ac}=\eta^{bd}R_{abcd},\qquad
R=\eta^{ac}R_{ac},\qquad
G_{ac}=R_{ac}-\frac{1}{2}\Pi_{ac}R.
\nom}
Using expression (\ref{11.1}) and formulas (\ref{2}), (\ref{3}) and (\ref{6})-(\ref{8}), we write
these expressions in terms of the field
$z^A(y)$. For example, the scalar curvature can be written as
\disn{12.1}{
R=\lks \Pi^{ac}\Pi^{bd}\ls\dd_a\dd_c z^A\rs w_{AB}\ls\dd_b\dd_d z^B\rs\rks_{cd}.
\nom}
Using this expression, we can try to construct the action for the field $z^A(y)$ whose equations of motion
would provide a correct description of the gravitational interaction.

\section{The theory action}
Because we assume that different surfaces $W$ do not interact, it is reasonable to write the action as an
integral over the space $Z$ of values of the function $z^A(y)$:
\disn{15}{
S=\int dz\; S_W(z),
\nom}
where $S_W(z)$ is the contribution of the surface $W$ with the given value of $z$ to the action (more precisely,
$S_W(z)dz$ is the contribution of surfaces corresponding to a small neighborhood of this point). Assuming
locality of the action and again introducing the temporary coordinates $x^\m$ on each surface $W$, we can write
$S_W(z)$ in the form
\disn{16}{
S_W(z)=\int d^4x \;\lc(x,z),
\nom}
where $\lc(x,z)$ is the quantity representing the scalar density with respect to coordinate transformations on
the surfaces.

In addition to $z^A(y)$, there can be other fields, matter fields, in $R^{1,N-1}$. We construct the theory
ensuring that events occurring on different surfaces are independent. All excitations including those for the
matter fields must propagate only along the surfaces. In the future, we can in principle consider variants of
the theory in which such an interaction is present but is sufficiently weak to not contradict observational
facts. Here, we restrict ourself to the case where such an interaction between surfaces is totally absent. We
assume that excitations of the matter fields propagate only along the surfaces if the action contains the
derivatives of the matter fields only in the directions along $W$ and the total action can be written as a sum
of actions on different surfaces.

Because our aim within this theory is to obtain a gravity theory on the surface $W$ close to Einstein's
General Relativity, it is logical to take the scalar density $\lc(x,z)$ in the form
\disn{17}{
\lc(x,z)=\dz\sqrt{-g}\ls-\frac{1}{2\ka} R+\lc_m\rs,
\nom}
where $\lc_m$ is the scalar quantity determining the matter field contribution to the action and $\ka$ is the
gravitational constant. We assume that this quantity contains differentiations only in the directions along
$W$. For example, we take the action $\lc_m$ for the scalar field in the form
\disn{18}{
\lc_m=\frac{1}{2}\ls \bar\dd_a\ff\rs\ls\bar\dd^a\ff\rs-V(\ff),
\nom}
where $V(\ff)$ is a potential and we use notation (\ref{13}). We note that reasonings analogous to formula (\ref{11})
result in rewriting expression (\ref{18}) in the standard form
\disn{18.1}{
\lc_m=\frac{1}{2}g^{\m\n}\ls\dd_\m\ff\rs\ls\dd_\n\ff\rs-V(\ff).
\nom}
As a result, theory action (\ref{15}) becomes
\disn{19}{
S=\dz\int dz\, d^4x\sqrt{-g}\ls-\frac{1}{2\ka}R+\lc_m\rs.
\nom}
In this integral, we pass from the curvilinear coordinates $\tilde y^a=\{x^\m,z^A\}$ (we assume that $\tilde y^a$
is the union of
the quantities $x^\m$ and $z^A$) to the coordinates $y^a$ related to $\tilde y^a$ by the formula
\disn{19.1}{
\tilde y^a(y)=\{x^\m(y),z^A(y)\}.
\nom}
We can show that the Jacobian of this transformation is given by
\disn{25a}{
J=\det\frac{\dd\tilde y^b}{\dd y^a}=\sqrt{\frac{w}{g\det\eta}}=\frac{\sqrt{|w|}}{\sqrt{-g}}.
\nom}
Using this formula, we can rewrite action (\ref{19}) as
\disn{27}{
S=\dz\int dy\, \sqrt{|w|}\ls-\frac{1}{2\ka}R+\lc_m\rs.
\nom}
If we use formula (\ref{12.1}) in this expression for the scalar curvature and use the formula of type (\ref{18}) for $\lc_m$,
then the coordinates $x^\m$ already drop out of these expressions. As a result, we do not need to introduce
any coordinate system on the surface $W$, and we have a form of writing the action that is natural from
the standpoint of considering the theory of the field $z^A(y)$ in the flat space-time $R^{1,N-1}$. We can say that
action (\ref{27}) is written in gauge-invariant terms with respect to the group of general covariant coordinate
transformations on four-dimensional manifolds.

It is interesting that the constructed action is not invariant under symmetry transformation (\ref{1}), which
was assumed to be physical. The quantities $R$ and $\lc_m$ are invariant (see the remark after formula (\ref{8})) but
the determinant $w$ is not invariant. Formulas (\ref{2}) and (\ref{3}) imply that under transformation (\ref{1}), a factor
of the form of the absolute value of the Jacobian of this transformation appears in the integrand of (\ref{27}).
We can easily obtain the same result if we use form (\ref{19}) of writing the action because transformation (\ref{1})
is a change of coordinates in the space $Z$ over which we integrate. But the factor arising in the action
depends only on $z^A$, not directly on $y^a$, and we can therefore move it outside the integration over $x^\m$ in
representation (\ref{19}). It therefore plays the role of the weight factor with which different noninteracting
surfaces contribute to the action and therefore does not affect the equations of motion. That this is in fact
the case becomes clear at the end of the next section. As a result, an interesting situation arises: the action
is not invariant with respect to some symmetry, but the equations of motion are.

\section{The equations of motion}
We find the equations of motion of the theory under consideration. Because the theory action was
initially constructed as the sum of contributions of each of the surfaces $W$, we logically expect that each
of the surfaces is governed by the same equation as if it would be alone, i.~e., by the Regge-Teitelboim
equation \cite{regge}. We obtain this result accurately. We can do this by directly varying action (\ref{27}) with respect
to the independent variable $z^a(y)$ and the matter fields. We then calculate in an explicitly gauge-invariant
way with respect to general covariant transformations because we do not introduce coordinate systems on
the surfaces $W$.
Thus derivation of the equations of motion was done by A.~Gromov
in bacalaurean thesis (department of High Energy
and Elementary Particles Physics, Physical Faculty, Saint Petersburg
State University).
But this direct derivation requires
cumbersome calculations; therefore, we here obtain the equations of motion in another way using the
possibility of introducing the coordinates $x^\m$ on the surfaces $W$ as an intermediate step and using the
results of the embedding theory.

Introducing the coordinates $x^\m(y)$, we write the action in form (\ref{19}). We can describe the surface $W$
corresponding to a definite value of $z^A$ by the embedding
function $y^a(x)=y^a(x,z)\mid_{z=const}=y^a(\tilde y)\mid_{z=const}$
where $y^a(\tilde y)$ is the function inverse to (\ref{19.1}). The metrics in action (\ref{19})
can be expressed via the embedding function by formula (\ref{v1}).

We find which variation of the embedding function $y^a(x)$ corresponds to an arbitrary small variation of
the independent variable $z^A(y)$. For an arbitrary point $\hat y^a$ (written in the Cartesian coordinates) of $R^{1,N-1}$,
we have the identity
\disn{30}{
y^a(x^\m(\hat y),z^A(\hat y))=\hat y^a,
\nom}
because $y^a(\tilde y)$ is the function inverse to (\ref{19.1}).
We pass from the field $z^A(y)$ to the result of its small variation,
\disn{30.1}{
z'^A(y)=z^A(y)+\de z^A(y).
\nom}
We assume that the function $x^\m(y)$ fixing the coordinates on $W$ simultaneously undergoes an arbitrary
small variation, thus transforming into the quantity ${x'}^\m(y)=x^\m(y)+\de x^\m(y)$. Then the embedding function
${y'}^a(x)$, which was changed because of the variation, satisfies the analogue of relation (\ref{30})
\disn{30.2}{
{y'}^a({x'}^\m(\hat y),{z'}^A(\hat y))=\hat y^a.
\nom}
Introducing the notation ${y'}^a(x)=y^a(x)+\de y^a(x)$, expanding the left-hand side of Eq.~(\ref{30.2}),
and using (\ref{30}), we obtain
\disn{31}{
\de y^a(x,z)=-\de z^A \frac{\dd}{\dd z^A} y^a(x,z)-\de x^\m\,\dd_\m y^a(x,z).
\nom}
The second term in the right-hand side of this formula is an arbitrary vector tangent to $W$. This arbitrariness
follows from the arbitrariness in choosing the coordinates $x^\m$; below, we show that it does not affect the
final form of the equations of motion.

We now consider how the matter fields respond to variation (\ref{30.1}). The field $\ff(y)$ as a function of $y^a$ is
unchanged, being an independent variable together with $z^A(y)$. But it enters action (\ref{19}) as a function of
$z^A$ and $x^\m$ in the form $\ff(x,z)=\ff(y(x,z))$ and therefore has the increment
\disn{32}{
\de\ff(x,z)=\ls\de z^A \frac{\dd}{\dd z^A} y^a(x,z)+\de x^\m\,\dd_\m y^a(x,z)\rs\dd_a\ff(y).
\nom}

We now use the results of the embedding theory. We know the variation of action (\ref{19}) of a single surface
with fixed $z^A$, i.~e., the variation of the expression in the integral over $z$ (see \cite{tmf07}, \cite{piyf06}) and can therefore write
\disn{33}{
\de S=\int\! dz\int\! d^4x\ls-\frac{\dz}{\ka}\sqrt{-g}\ls G^{\m\n}-\ka\, T^{\m\n}\rs b^a_{\m\n}\de y_a(x,z)+
\frac{\de S}{\de\ff(x,z)}\de\ff(x,z)\rs,
\nom}
where $G^{\m\n}$ is the Einstein tensor and $T^{\m\n}$ is the matter energy-momentum tensor calculated standardly,
i.~e., by varying the contribution of a single surface to the action with respect to the metric. We note
that it does not coincide with the energy-momentum tensor of matter fields considered as some fields in
the flat ambient space $R^{1,N-1}$. Our reasonings before formula (\ref{32}) explain the origin of the last term in
expression (\ref{33}). Substituting (\ref{31}) and (\ref{32}) in (\ref{33}) and using the fact that the quantity $b^a_{\m\n}$ with respect
to the index $a$ is orthogonal to any tangent vector, we obtain the final expression for the action variation
under an arbitrary variation of the independent variable $z^A(y)$:
\disn{33.1}{
\de S=\int\! dz\int\! d^4x\Biggl(\frac{\dz}{\ka}\sqrt{-g}\ls G^{\m\n}-\ka\, T^{\m\n}\rs b^a_{\m\n}\frac{\dd y_a(x,z)}{\dd z^A}\,\de z^A+\nss+
\frac{\de S}{\de\ff(x,z)}\ls\de z^A \frac{\dd}{\dd z^A} y^a(x,z)+\de x^\m\,\dd_\m y^a(x,z)\rs\dd_a\ff(y)\Biggr).
\nom}
We note that the completely arbitrary quantity $\de x^\m$ related to choosing the coordinates enters this expression.
But it does not enter the final form of the equations of motion. Indeed, in addition to the variation of
the action with respect to the field $z^A(y)$, we must also consider the variation with respect to the matter
field $\ff(y)$ when deriving the complete set of equations of motion. The resulting matter equation of motion
can be written in the general case as
\disn{34}{
\frac{\de S}{\de\ff(x,z)}=0,
\nom}
while it takes the standard form for specific theory (\ref{18}), (\ref{18.1}),
\disn{34.1}{
g^{\m\n}D_\m \dd_\n\ff+V'(\ff)=0,
\nom}
where $D_\m$ is the standard covariant derivative in the coordinate system $x^\m$. Taking Eq. (\ref{34}) and the
above orthogonality of the quantity $b^a_{\m\n}$ with respect to the index $a$ into account, we find that equating
variation (\ref{33.1}) to zero at an arbitrary $\de z^A$ results in
\disn{35}{
\ls G^{\m\n}-\ka\, T^{\m\n}\rs b^a_{\m\n}=0,
\nom}
i.~e., in the Regge-Teitelboim equation known from the embedding theory.

The obtained equations of motion (\ref{34.1}), (\ref{35}) are now written in the form related to the choice of the
coordinates $x^\m$ on the surfaces $W$ because we use the coordinate systems in the derivation. But we can also
write them in the form of equations that do not involve the coordinates $x^\m$ using the formalism in \cite{tmf07}, \cite{piyf06}
and formulas in Sec. 2. We can rewrite Eq. (\ref{34.1}) in the form
\disn{36}{
\bar\dd_a\bar\dd^a\ff+V'(\ff)=0,
\nom}
where we use notation (\ref{13}). Regge-Teitelboim equation (\ref{35}) in the coordinate-free form is
\disn{37}{
\ls G^{cd}-\ka\, T^{cd}\rs b^a_{cd}=0,
\nom}
where $G^{cd}$ and $b^a_{cd}$ are defined by (\ref{12}), (\ref{11.1}), and (\ref{13.1})
and where $T^{cd}$ results from "transferring" the standard
matter energy-momentum tensor from $W$ to the ambient space. For example, this tensor for theory (\ref{18}) is
\disn{38}{
T_{cd}=e^\m_c e^\n_d\, T_{\m\n}=\ls\bar\dd_c\ff\rs\ls\bar\dd_d\ff\rs-
\Pi_{cd}\ls\frac{1}{2}\ls\bar\dd_a\ff\rs\ls\bar\dd^a\ff\rs-V(\ff)\rs.
\nom}
It is interesting that the obtained equations of motion are invariant under transformation (\ref{1}), which
is easily seen from formulas (\ref{8.2}), (\ref{8.3})-(\ref{12}).
We can therefore consider that the symmetry under transformation (\ref{1}) is physical even though
action (\ref{27}) is not invariant under these transformations (we already
discussed this situation at the end of Sec. 3).

\vskip 0.5em
{\bf Acknowledgments.}
The author thanks the organizers of the Third International Conference "Models
of Quantum Field Theory", dedicated to the 70th birthday of Aleksandr Nikolaevich Vasiliev.


\begin{thebibliography}{1}

\bibitem{carlip}
S.~Carlip, \emph{Rept. Prog. Phys.}, v.~64, pp.~885--942, 2001,
  arXiv:gr-qc/0108040.

\bibitem{regge}
T.~Regge and C.~Teitelboim, "General relativity \`a la string: a progress
  report", in \emph{Proceedings of the First Marcel Grossmann Meeting,
  Trieste, Italy, 1975}, ed. R.~Ruffini,  1977, pp.~77--88.

\bibitem{sbshk05}
S.A.~Paston and V.A.~Franke, "The gravity as a theory of embedding of space-time
  into the flat space of higher dimensions [in Russian]", in \emph{Proceedings
  of the 15 International V.A. Fock school for advances of physics 2005},
  ed. V.~Novozhilov, Publishing house of
  St.Petersburg State University, 2006, p.~34.

\bibitem{tmf07}
S.A.~Paston and V.A.~Franke, \emph{Theoretical and mathematical physics}, v. 153,
  no.~2, pp. 1582--1596, 2007, arXiv:0711.0576.

\bibitem{gane}
M.~Janet, \emph{Ann. Soc. Math. Pol.}, v.~5, pp.~38--43, 1926.

\bibitem{kart}
E.~Kartan, \emph{Ann. Soc. Pol. Math.}, v.~6, pp.~1--7, 1927.

\bibitem{kobno2}
S.~Kobayashi and K.~Nomizu, \emph{Foundations of Differential Geometry}.\hskip
  1em plus 0.5em minus 0.4em\relax Wiley, 1969, v.~2.

\bibitem{ijtp10}
S.A.~Paston and A.N.~Semenova, \emph{Int. J. Theor. Phys.}, v.~49, no.~11, pp.~2648--2658, 2010, arXiv:1003.0172.

\bibitem{piyf06}
S.A.~Paston and V.A.~Franke, "Einstein's general relativity as a theory of 4d
  surface in a flat 10d space [in Russian]", in \emph{Collection of Materials
  of XLI and XLII Winter Schools of SPINP "Physics of Atomic Nuclei and
  Elementary Particles"}. St.Petersburg
  Inst. Nucl. Phys., 2008, pp.~231--275.

\end{thebibliography}

\end{document}